\documentclass[aps,prb,twocolumn,unsortedaddress,showpacs]{revtex4-1}

\usepackage{graphicx}
\usepackage{bm}

\begin{document}

\title{Upper limit of supersoldity in solid helium }

\author{Duk Y. Kim}

\author{Moses H. W. Chan}
\email[]{chan@phys.psu.edu}

\affiliation{Department of Physics, Pennsylvania State University, University Park, Pennsylvania 16802, USA}

\date{\today}

\begin{abstract}
The resonant period drop observed at low temperatures in torsional oscillators containing solid helium had been interpreted as a signature of supersolid. However, it was found that the shear modulus increase found in solid helium at the same low temperature could also decrease the resonant period of the torsional oscillator. We report the results of a study in two different torsional oscillators that were specially designed to minimize the shear modulus effect and maximize any possible supersolid response. We were able to place an upper limit on the nonclassical rotational inertia or supersolid fraction of $4\times10^{-6}$. Moreover, we have repeated an earlier experiment on hcp $^3$He solid, which shows similar low temperature stiffening as in hcp $^4$He. We found that the small drop of the resonant period measured in the hcp $^3$He sample is comparable in size to that found in the hcp $^4$He samples. These results strongly suggest that the resonant period drop reported in most torsional oscillator studies in the last decade is primarily a consequence of the shear modulus stiffening effect.
\end{abstract}

\maketitle

\section{Introduction}

Evidences of possible nonclassical or missing rotational inertia in solid helium were first reported by Kim and Chan in their torsional oscillator (TO) experiments in \,2004.\cite{NatureKC,ScienceKC} When the torsion cell filled with solid helium was cooled below \,200 mK, a drop in the resonant period of the TO was found. The magnitude of the period drop increases with decreasing temperature and plateaus below 60 mK. The magnitude of the period drop diminishes with the oscillation velocity of the torsion cell for velocities exceeding approximately 10 micrometer per second. The drop in the resonant period has been interpreted as a signature of nonclassical rotational inertia (NCRI) or superfluidity in solid helium and the dependence on the oscillation velocity as a superfluid critical velocity effect. The NCRI, i.e., the superfluid fraction, of solid helium can be calculated by dividing the period drop by the mass loading, i.e., the resultant increase in the resonant period of the TO due to the introduction of the solid helium sample. The observations of the TO anomaly have since been reported in over thirty other TO experiments carried out in eleven laboratories.\cite{Shirahama,Kojima07,Reppy07,Kubota,Hunt,PG2010,NMRTO,Eyal,Manchester,Fefferman,Nichols} 

An important challenge in the study of supersolidity is to understand the causal relation between the period drop observed in TO experiments and the observation of an increase in the shear modulus of solid helium. The shear modulus of polycrystalline solid helium increases up to 20\% below \,200 mK and it tracks the TO anomaly with identical temperature and $^3$He concentration dependences.\cite{DayBeamish} In single crystal samples the shear modulus can increase as much as 80\%.\cite{Rojas10} The mechanism behind the shear modulus increase is the binding of dislocation lines to $^3$He impurities. Since solid helium is a constituent of the TO, an increase in the shear modulus of solid helium will stiffen the TO and causes the resonant period to drop, thus giving an apparent NCRI. For a perfectly rigid TO with a simple geometry, this shear modulus effect on the resonant period can be calculated analytically. For a TO oscillating at 1 kHz and containing an isotropic solid helium sample in the shape of a cylinder with 1 cm in diameter and height, the 20\% increase in the shear modulus of solid helium results in an apparent NCRI of approximately $10^{-4}$. If the solid helium sample is changed to an annulus shape, the apparent NCRI can be reduced by at least one order of magnitude.\cite{Maris11,ReppySM} For an actual TO with a more complicated geometry, this shear modulus effect can be calculated numerically by the finite element method (FEM). While a few experiments found period drops that are only a few times larger than the values calculated for the increase in the shear modulus of 20\%\cite{LongPath}, the great majority of the TO experiments found period drops that are two or even three orders of magnitude larger. These discrepancies led to the argument that the shear modulus effect cannot account completely the observed period drops and leave the issue of supersolidity open. Indeed, two very recent experiments employing TO with two resonant modes reported NCRI of $1.3\times10^{-3}$,\cite{Nichols} and $1.2\times10^{-4}$.\cite{Reppy14}

However, there are three mechanisms that can greatly amplify the shear modulus effect. If a TO is  not perfectly rigid, i.e., one part of the torsion cell can move with respect to the other parts, solid helium may act as a glue in coupling the oscillatory motion of the different parts of the torsion cell. In such a situation, the stiffening of solid helium will have a much larger effect on the resonant period of the TO. Most TOs assembled with epoxy possibly have this problem because helium can penetrate an imperfect epoxy joint and this thin solid helium layer contributes to the gluing function of the epoxy in coupling the different parts of the TO in oscillatory motion. FEM simulations find that the effectiveness in changing the resonant period of the TO increases when the thickness of the solid helium layer is reduced. In some TO, the torsion rod is connected to the main body of the torsion cell via a thin metal plate that is in contact with the solid helium sample. In this case, the main body of the TO and the torsion rod will oscillate with a difference phase and the thin metal plate and the solid helium adhering to the plate together can be considered as an additional ``spring'' of such a compound oscillator. The shear modulus effects of helium in these TOs are naturally larger than the ideal case of a completely rigid TO.\cite{Maris12,Vycor2012} Some TOs have  hollow torsion rods that serve as  fill lines for the helium samples. If the ratio of the outer and inner diameters of such a hollow torsion rod is not sufficiently large, the solid helium inside the torsion rod contributes to the spring constant of the torsion rod and induces a measurable period drop when it stiffens.\cite{BeamishRod}

TOs with rigid constructions do show smaller period drops that are consistent with the shear modulus stiffening interpretation. In our recent experiment to search for a path length dependence in the possible NCRI, TOs in the shape of toroids and self-connected loops were fabricated entirely out of metallic (stainless steel or Cu-Ni) capillaries and tubes. The joints in these TOs were brazed together with silver solder.\cite{LongPath} The path lengths of these samples range from 6 to 100 cm. No evidence of period drop were found in two TOs within the resolution (0.1 ns) of the measurements. In the other two TOs, very small period drops (0.47 ns and 0.55 ns) were found. No evidence of any path length dependence was found. If we assume there is no shear modulus effect, the measured period drops correspond to NCRI of $3\times10^{-5}$ and $4\times10^{-5}$. For the two TOs finding no measurable period drop, the resolution in the period reading (0.1 ns) translates to an upper limit in NCRI of $7\times10^{-5}$ and $4\times10^{-5}$. The small period drops found in these TOs are most likely a consequence of their rigid constructions. In addition, the small cross section of the tube and the capillary forming the sample space also substantially reduces the shear modulus stiffening effect. Nevertheless, FEM simulations showed that the observed period drops are 6 and 3 times larger than that calculated for a 20\% increase in the shear modulus reported for polycrystalline solid samples. There are four possible interpretations of these results: (1) There is no NCRI and the TOs are perfectly rigid. The larger than expected period drops indicate that the shear moduli increase far more than 20\% at low temperatures because the solid samples are single crystals rather than polycrystalline. This explanation is unlikely to be correct since we did not take any measures to grow single crystal samples. (2) There is no NCRI and the shear modulus effect was amplified because the TOs are not perfectly rigid. (3) There is no NCRI and the TOs are sufficiently rigid but the FEM simulations underestimated the effect of the shear modulus increase. (4) The observed period drop is a consequence of a shear modulus increase and a small NCRI of approximately $3\times10^{-5}$.

The shear modulus stiffening effect, owing to its origin in the binding of the dislocation network to $^3$He impurities, is not relevant for solid helium confined in porous Vycor glass since its porous structure is too restrictive to accommodate a dislocation network. It is therefore tempting to argue that the \,2004 Kim and Chan paper\cite{NatureKC} that reported a period drop in the solid helium confined in Vycor glass is a signature of NCRI free from the shear modulus effect. However, the construction of the \,2004 Vycor TO makes it unavoidable to have a thin bulk solid helium layer inside the torsion cell. In 2012, a new Vycor TO completely free from any bulk solid was built. With the new TO, no evidence of any period drop down to experimental resolution of $2\times10^{-5}$ in NCRI was found.\cite{Vycor2012} In a separate experiment, a Vycor TO with two resonant modes was employed to distinguish superfluid-like signals from other effects. It was concluded that most of the observed period drops are frequency dependent and not compatible with the NCRI interpretation.\cite{MiReppy}

The results from the rigid TOs and Vycor TOs strongly suggest that at least a large fraction of the period drops seen in most TO experiments are due to the shear modulus stiffening effect. NCRI, if it exists, is at most a few parts in $10^{-5}$.\cite{Review2013,*Review2013Err} In this paper we report our effort in nailing down the issue on the existence of supersolidity in bulk solid $^4$He. We built two particularly rigid TOs that are optimized to enhance any NCRI signal, if it exists, while minimizing the effect of shear modulus stiffening. In addition, we carried out new measurements on solid $^3$He and $^4$He to revisit the quantum statistics effect reported in 2009.\cite{West}

\section{Measurement on solid $^3$H\MakeLowercase{e}}

\subsection{Background}

It was shown in the 2009 experiment that the shear modulus of hcp $^3$He solid shows an increase similar to that found in hcp $^4$He. Such an increase was not seen in bcc $^3$He.\cite{West,SyshchenkoBeamish}. This dependence on the crystal structure confirms the dislocation interpretation of the modulus changes because the structure and mobility of the dislocation network should depend strongly on the crystal structure but not on the quantum statistics. However, in the companion torsional oscillator study, no period decrease at low temperatures was observed in hcp $^3$He. When the same TO was filled with hcp $^4$He solid, a period drop of 0.5 ns was seen. While this period drop is small, it is 30 times larger than the calculated value for a 20\% increase in the shear modulus. The oscillator used was believed to be very rigid since it was made by welding two pieces of beryllium copper to form an annulus sample space. Therefore, the result of this 2009 experiment supports the interpretation that the observed period drop is a signature of NCRI since it is seen in a Boson and not in a Fermion solid. However, there is one concern about the experiment in 2009. Specifically, for the set of measurements on solid $^3$He samples (of both hcp and bcc structures), the resonant period of the empty oscillator is found to increase by 3 ns when the temperature is reduced from 600 to 20 mK. Over the same temperature range, the $Q$ factor changes by a factor of 4. Such changes were not seen for measurements on solid $^4$He samples that took place prior to the measurements on $^3$He samples.\footnote{The TO was warmed up to room temperature and flushed  with $^3$He after the set of measurements on $^4$He. When it was cooled down and pressurized with $^3$He, a leak was found in the TO. The leak at the weld joint was repaired with epoxy for the set of measurements on solid $^3$He.} The resonant period and the $Q$ were found to be essentially temperature independent in the low temperature limit similar to those observed in most torsional oscillators. The change in period was found to be less than 0.1 ns and in $Q$ to be less than 15\% between 25 and 450 mK. This anomalous behavior in the solid $^3$He data set drove us to repeat the experiment to resolve the mystery.

\subsection{Experimental Details}

The 2009 torsional oscillator was used in the new measurements. New sample fill lines were installed on the cryostat and the TO was flushed with nitrogen and $^3$He at room temperature to eliminate any possible contamination of the $^3$He solids with $^4$He impurities. The same source of $^3$He gas with 1.35 ppm $^4$He as in 2009 was used to make the $^3$He solids. After the measurements on $^3$He samples, we flushed the TO with $^4$He at 30 K and made measurements on pure $^4$He solids with 0.3 ppm $^3$He. We made measurements on empty cell backgrounds before and after the measurements on each and every solid samples we studied and found the results to be consistent without showing any anomalous change in the period and $Q$ between 30 and 700 mK (Fig. 1). The resonant period is 2.94 ms (340 Hz in frequency), identical to that found in 2009. In order to grow the hcp $^3$He solid sample, the TO was pressurized with 150 bar liquid $^3$He at 3.4 K and cooled down. Freezing commenced below 3.4 K and completed at 2.8 K. Based on the phase diagram\cite{Dobbs} of $^3$He, the pressure of the sample is estimated to be 120 bar, well inside the hcp phase. The bcc $^3$He and the hcp $^4$He samples were prepared with similar procedures at lower pressures.

\subsection{Results}

\begin{figure}[tb]
    \centerline{\includegraphics[width=1\columnwidth]{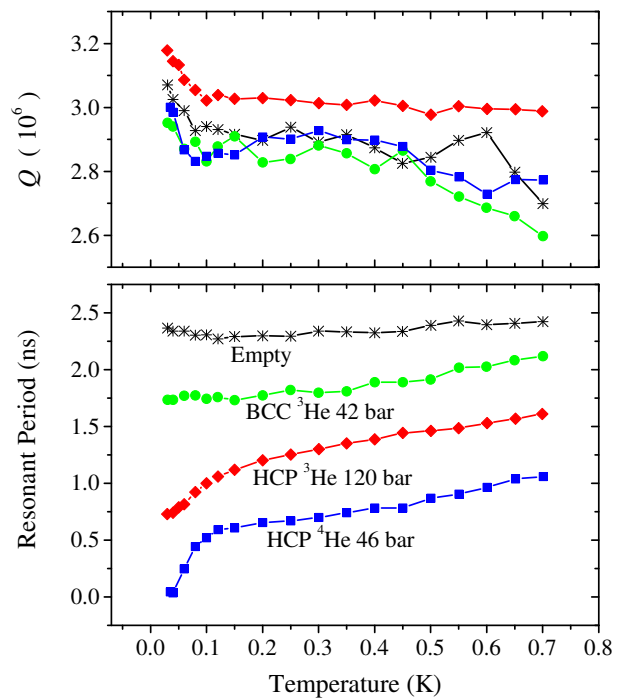}}
    \caption{(Color online) New measurements of the torsional oscillator (used in Ref. \onlinecite{West}) containing different solid helium samples. The period data are shifted for easy comparison. The mechanical $Q$ data of the different samples are shown in the top panel with the same symbols as the period. The oscillation velocities for all the data are between 10 and 20~$\mu$m/s.}  
    \label{Fig1}
\end{figure}

\begin{figure}[tb]
    \centerline{\includegraphics[width=1\columnwidth]{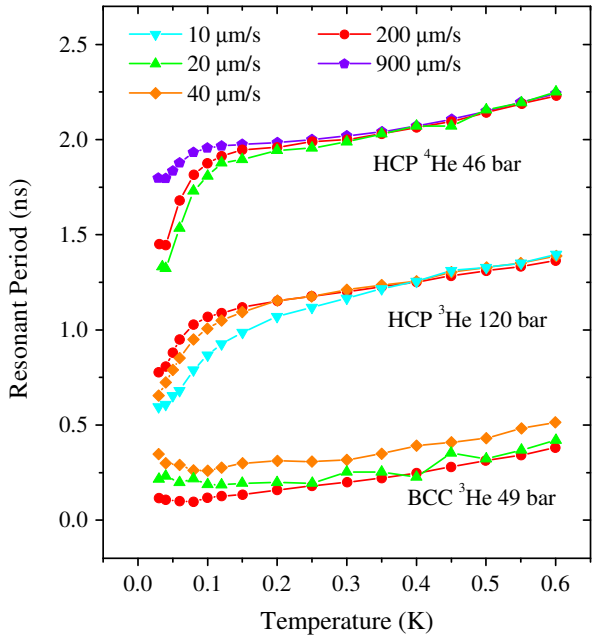}}
    \caption{(Color online) Resonant periods with different oscillation velocities. The data sets for different samples are shifted for easy display. Data taken with different oscillation speeds on the same sample were not shifted with respect to each other.}
    \label{Fig2}
\end{figure}

Figure 1 shows the resonant periods of the oscillator with different solid samples. The resonant period of the empty oscillator showed a slight decrease, approximately 0.2 ns, as the temperature was reduced from 700 to 30 mK. The anomalous and large (3 ns) increase of the resonant period with decreasing temperature reported in 2009 was not observed in current measurements. A period drop of 0.7 ns was found in the hcp $^4$He below 200 mK. This value is nearly the same as that reported in 2009. No low temperature period drop was observed in bcc $^3$He, also consistent with that reported in 2009. However, unlike the previous results, we found a period drop in the hcp $^3$He solid. The magnitude of the period drop is approximately 0.6 ns, nearly the same as that found in the hcp $^4$He solid. The period drop in hcp $^3$He commences at a slightly higher temperature than that found in hcp $^4$He. This is probably because the concentration of $^4$He impurities in the $^3$He solid (1.35 ppm) is higher than that of $^3$He impurities (0.3 ppm) in the $^4$He solid.\cite{DayBeamish,SyshchenkoBeamish} Figure 2 shows the resonant periods of the TO with different oscillation velocities. Similar to $^4$He samples, the period drop in hcp $^3$He was suppressed by high oscillation velocities. It was shown that the shear modulus increase in the hcp $^3$He solid was also suppressed by high strains.\cite{West} Therefore, it is very clear that the period drop observed in hcp solid $^3$He at low temperature has the same physical origin as in hcp $^4$He. Specifically it is the consequence of the shear modulus stiffening brought about by the binding of dislocation lines to isotopic impurities in the solid. 

In a few sets of the measurement with $^4$He samples, the mechanical $Q$ data showed dissipation peaks near 100 mK. Indeed, it was difficult to get stable $Q$ measurements in this TO at low velocities. The ring down time and hence the $Q$ value of this oscillator was found to be higher than most other TOs. However, both the resonant period and the amplitude of the oscillation were not particularly stable. It appears this TO may be exceptionally susceptible to external noise and we suspect the $^3$He solids data sets including the anomalous empty cell results reported in 2009 may be plagued by this instrumental problem.

\section{Rigid Torsional Oscillators}

\subsection{Design}

In order to further clarify the issue of supersolidity and set an upper limit for any possible NCRI, we designed new TOs that are particularly insensitive to shear modulus change of the solid helium sample. Figure 3 shows the drawing of the two different TOs used in this effort. The common feature of the two oscillator is that each has an annulus sample space formed between a heavy metal cylinder in the middle and a thin outer wall. TO~A was made by hollowing out an annulus sample space from a brass cylinder. The annulus space is then sealed by soldering a small cap. TO~B was made by assembling a stainless steel tube on a solid metal cylinder, which was machined to form an annulus sample space. By making the torsional cell with a single solid piece of metal and attaching it to the torsion rod rigidly with screws, the oscillatory motion of the torsion cell is driven solely by straining the torsion rod without straining the cell body. Although the center of the oscillator is heavy, it does not seriously degrade the sensitivity in measuring the rotational inertia contribution of the helium sample because the rotational inertia of a cylindrical shell scales as the fourth power of the radius from the axis of rotation. In contrast to the heavy center, we made the outer wall of the annulus very thin. By placing solid helium in the outermost part of the oscillator, we can also minimize the shear modulus effect of the solid helium on the resonant period of the oscillator. Our design maximizes the rotational inertia contribution of the solid helium sample but makes the TO insensitive to the change in the modulus of the solid helium sample. Our design also removes any extraneous crevices in the sample space that may house a thin solid helium layer and lead to unpredicted resonant period changes.

\begin{figure}[tb]
    \centerline{\includegraphics[width=1\columnwidth]{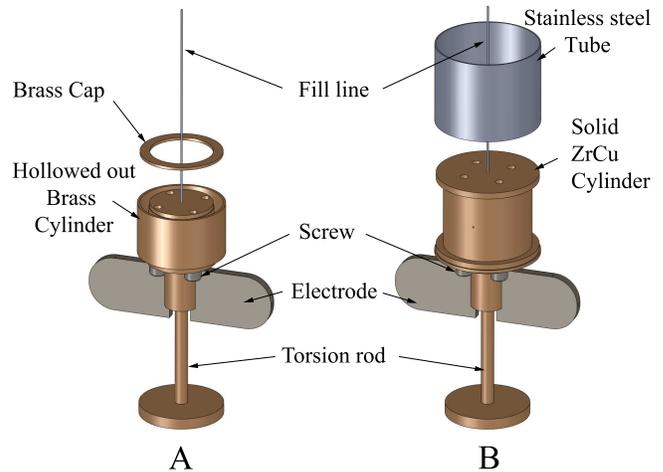}}  
    \caption{(Color online) Drawing of torsional oscillators designed to minimize the shear modulus effect. The oscillators have massive metals in the center and the annulus sample spaces are placed in the outermost part of the oscillators.}  
    \label{Fig3}
\end{figure}

\subsection{Experimental Details}

The annulus sample space of TO~A for solid helium has dimensions of 14.9 mm in outer diameter, 1.6 mm in width, and 7.6 mm in height. The outer wall of the annulus sample space is 0.5 mm thick. TO~B is made by silver soldering a thin stainless steel tube onto a solid zirconium copper cylinder. The annulus sample space of 18.5 mm in outer diameter, 1.3 mm in width, and 12.9 mm in height is formed between the stainless steel tube and the zirconium copper cylinder (Fig. 3). The thickness of the stainless steel wall is 0.13 mm. In both TOs, the copper nickel fill lines (outer diameter 0.3 mm and inner diameter 0.1 mm) for helium were connected into the annulus sample space through the center of the cylinder bodies. The torsion cells were rigidly attached by screws to the same torsion rod used in previous studies.\cite{LongPath,Vycor2012} In the viewpoint of the rigidity mentioned above, TO~B is more carefully designed to effectively exclude the effect from the shear modulus change of the solid helium. TO~B has a massive metal block in the center and an annulus sample space with a larger diameter and a thinner width. The Finite Element Method (FEM) simulation indicated TO~B is expected to have a factor of three smaller effect due to the shear modulus increase of the solid helium than TO~A. The resonant period of TO~A is 1.47 ms (679 Hz in frequency). The expected resonant period change due to the loading of the solid helium sample, mass loading, is approximately 9\,000 ns. The resonant period of TO~B is 2.56 ms (390 Hz in frequency) and the expected mass loading is approximately 14\,000 ns. According to FEM simulations, the expected resonant period change for a 20\% increase in the shear modulus increase of solid helium for TO~A and TO~B are respectively 0.06 ns and 0.02 ns.

The standard TO technique was used in the measurements. TOs are kept oscillating at the resonant periods by constant driving ac voltages applied to one of the two electrodes. The oscillation amplitudes and velocities are measured with the induced currents on the other electrodes. The mechanical $Q$ of a TO is proportional to and can be calculated from the amplitude of the oscillation. The blocked capillary method was used in growing all the solid helium samples inside the TOs. The temperature was monitored with a thermometer attached on each TOs. The pressure of the solid helium sample was determined from the temperature at which the solidification of the sample is completed.

\subsection{Result of TO~A}

\begin{figure}[tb]
    \centerline{\includegraphics[width=1\columnwidth]{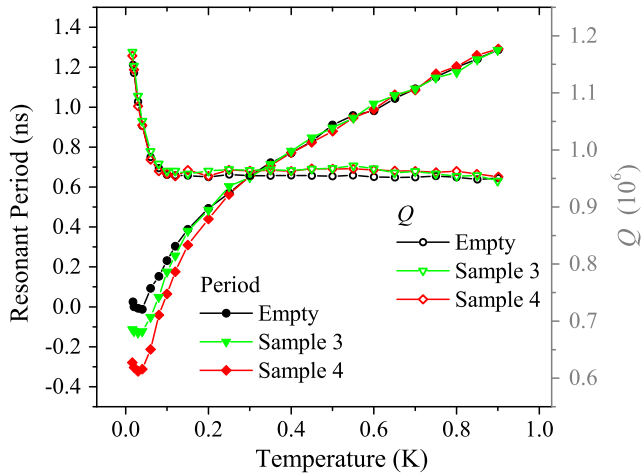}}
    \caption{(Color online) The resonant period and $Q$ data of TO~A. Sample 3 and 4 are the samples that showed the minimum and the maximum period drop at low temperatures among the 10 samples studied. The resonant periods for each samples are shifted to collapse above 300 mK. The mechanical $Q$ values are not shifted. The oscillation velocities for all the data are between 15 and 19 $\mu$m/s. The pressures of sample 3 and 4 are both 50 bars.}  
    \label{Fig4}
\end{figure}

Figure 4 shows typical results of the measurement with TO~A. The resonant periods of the empty oscillator and the oscillator filled with solid helium samples are found to be exactly parallel to each other above 300 mK. (The data shown in the figure are shifted by the mass loading values.) Below 300 mK, the resonant periods of the solid samples have lower values than those of the empty oscillator. Ten solid helium samples with the natural isotopic purity (0.3 ppm $^3$He) and the pressure range from 40 to 60 bars were tested. The period drop ranges between 0.12 and 0.32 ns and the average value is 0.22 ($\pm$0.06) ns. (Only the results of sample 3 and 4 are shown in the figure.) This period drop is a factor of 3.6 larger than the 0.06 ns value calculated for a 20\% increase in shear modulus. Since the average mass loading was 9\,200 ns, NCRI would be $2.4\times10^{-5}$ if we attribute the period drop entirely to supersolidity. The mechanical $Q$ increases below 100 mK and this is the typical behavior of many torsional oscillators. The $Q$ of the empty oscillator and those with all solid helium samples show the same temperature dependences with no sign of solid helium related dissipation at low temperatures.

\subsection{Result of TO~B}

\begin{figure}[tb]
    \centerline{\includegraphics[width=1\columnwidth]{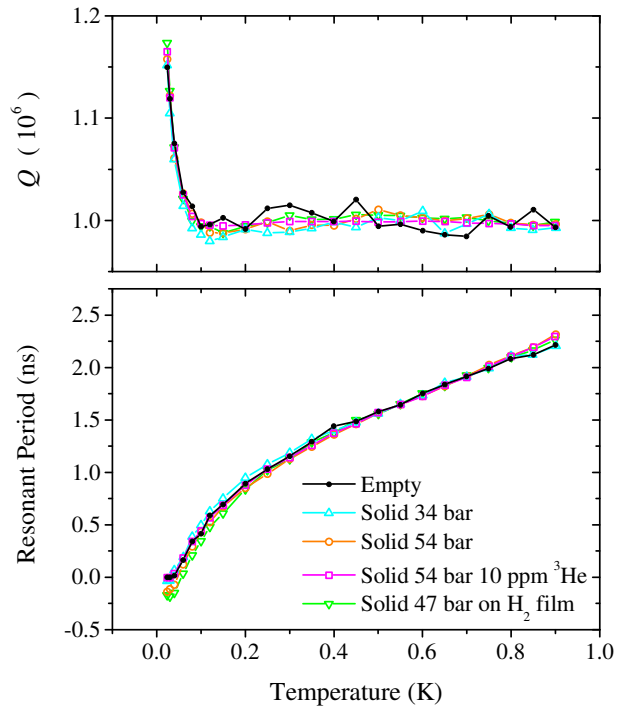}}
    \caption{(Color online) The resonant period and $Q$ data of TO~B. The resonant periods for each samples are shifted to collapse above 300 mK. The mechanical $Q$ values of the different samples are shown in the top panel with the same symbols as the period. The oscillation velocities for all the data are between 11 and 14 $\mu$m/s.}  
    \label{Fig5}
\end{figure}

Figure 5 shows the measurements of TO~B when it is empty and filled with different solid helium samples. Eight solid samples with the pressure range from 34 to 54 bars were measured. The resonant periods of the empty oscillator and the oscillator filled with solid helium samples are seen to be nearly parallel to each other over the entire temperature range of the measurements. If we shift the measured period readings of the 8 samples above 300 mK to collapse to the empty cell curve, the period drops of these samples in the low temperature limit of 30 mK range from 0.003 ns to 0.12 ns with the average value of 0.06 ($\pm$0.05) ns. This 0.06 ns value is close to the resolution of the measurement and the ``scatter'' in the different samples. Among the 8 solid samples, two samples were grown with 10 ppm $^3$He impurities and others with natural purity helium (0.3 ppm $^3$He). Nevertheless, the average period drop (0.06 ns) is a factor of three larger than the calculated value of 0.02 ns for a 20\% shear modulus effect. Since the average mass loading for the total samples is 14\,400 ns, NCRI would be $4\times10^{-6}$ if we attribute the period drop entirely to supersolidity. Again, the $Q$ of the empty oscillator and those with solid helium coincide over whole temperature range and no sign of dissipation is observed. 

We also made measurements on solid $^4$He (0.3 ppm $^3$He impurity) grown after preplating the inner walls of the torsion cell with several molecular layers of hydrogen and nitrogen to search for a possible surface effect. No significant difference was found in these samples as compared with solid samples grown in the bare sample cell. The period drop found in the sample with hydrogen preplating is 0.17 ns, marginally larger than other samples~(Fig. 5).

\section{Discussions}

We have made measurements with TOs that were optimized to minimize the shear modulus effect and simultaneously maximize their sensitivity for detecting NCRI. These measurements allow us to put an upper limit on NCRI of no more than $4\times10^{-6}$. The period drop of 0.22 ns was found with TO~A and 0.06 ns with TO~B with respective mass loadings of 9\,200 ns and 14\,400 ns. Both results are 3-4 times larger than the calculated shear modulus effect. It is interesting that in our earlier measurements with toroidal TOs the measured period drops were 3-6 times larger than the calculated values.\cite{LongPath} There are two likely explanations for this interesting ``systematic'' discrepancy seen in TOs of different designs with expected shear modulus effects that differ by an order of magnitude. The first one is that the solid samples in all the TOs are single crystal samples exhibiting the increase in shear modulus that is several times larger than the 20\% value we have used in the simulation. The second one is that the FEM simulations we have carried out underestimated the effect of the shear modulus increase in real TOs. As noted above, we think the second explanation is more likely to be correct. 

The new measurements on hcp $^3$He solid provided further evidence that the observed period drop is not a consequence of NCRI. The period drops measured with both hcp $^3$He and $^4$He are much larger than the calculated shear modulus effect assuming a rigid TO. It appears that this TO is not as rigid as we have assumed. We think that in the process of welding the TO, porous open spaces between the joints, accessible for helium, probably have opened up and rendered the TO much more sensitive to the change in the shear modulus of solid helium. The fact that a leak developed upon pressurizing the torsion cell with helium as reported in the 2009 paper supports this scenario.\cite{West}

The results reported here suggest that there is to date no TO evidence of supersolidity in bulk solid helium. NCRI, if it exists, is smaller than $4\times10^{-6}$ according to TO experiments. For solid helium confined in porous Vycor,\cite{Vycor2012} the upper limit of NCRI was found to be $2\times10^{-5}$.

\begin{acknowledgments}
We thank John Beamish for useful discussions and providing the high purity $^3$He gas. Support for this experiment was provided by NSF Grant No. DMR1103159.
\end{acknowledgments}

\bibliography{RigidTO}
\end{document}